# Mitochondrial $Ca^{2+}$ uptake in skeletal muscle health and disease

Jingsong Zhou*, Kamal Dhakal & Jianxun Yi

*Kansas City University of Medicine and Bioscience, Dybedal Research Center, Kansas City MO 64106, USA*

Muscle uses $Ca^{2+}$ as a messenger to control contraction and relies on ATP to maintain the intracellular $Ca^{2+}$ homeostasis. Mitochondria are the major sub-cellular organelle of ATP production. With a negative inner membrane potential, mitochondria take up $Ca^{2+}$ from their surroundings, a process called mitochondrial $Ca^{2+}$ uptake. Under physiological conditions, $Ca^{2+}$ uptake into mitochondria promotes ATP production. Excessive uptake causes mitochondrial $Ca^{2+}$ overload, which activates downstream adverse responses leading to cell dysfunction. Moreover, mitochondrial $Ca^{2+}$ uptake could shape spatio-temporal patterns of intracellular $Ca^{2+}$ signaling. Malfunction of mitochondrial $Ca^{2+}$ uptake is implicated in muscle degeneration. Unlike non-excitable cells, mitochondria in muscle cells experience dramatic changes of intracellular $Ca^{2+}$ levels. Besides the sudden elevation of $Ca^{2+}$ level induced by action potentials, $Ca^{2+}$ transients in muscle cells can be as short as a few milliseconds during a single twitch or as long as minutes during tetanic contraction, which raises the question whether mitochondrial $Ca^{2+}$ uptake is fast and big enough to shape intracellular $Ca^{2+}$ signaling during excitation-contraction coupling and creates technical challenges for quantification of the dynamic changes of $Ca^{2+}$ inside mitochondria. This review focuses on characterization of mitochondrial $Ca^{2+}$ uptake in skeletal muscle and its role in muscle physiology and diseases.

**skeletal muscle, mitochondria, $Ca^{2+}$**

## INTRODUCTION

ATP is the major currency of energy for sustaining life and is mostly produced in mitochondria. At the expense of other nutrient substrates and oxygen, mitochondria produce ATP that can be exchanged instantly whenever intracellular energy is required (Knowles, 1980). As described in the historical review by O'Rourke (O'Rourke, 2010), mitochondria, when initially discovered by Richard Altmann in 1890, were called "bioplast", meaning "life germs". The word "mitochondria" was given by Carld Benda in 1898. For decades mitochondria were studied as the power house of cell, and soon it was realized that $Ca^{2+}$ entry into mitochondria is required to stimulate the Krebs cycle and electron transport chain activity that result in enhanced ATP synthesis inside mitochondria (Balaban, 2002; Carafoli, 2014; Denton et al., 1980; Drago et al., 2011).

$Ca^{2+}$ is fundamental to normal cellular function. Cells possess specialized mechanisms to ensure a tightly controlled intracellular $Ca^{2+}$ level. These mechanisms involve complex interplay between intracellular $Ca^{2+}$ storage, buffering and $Ca^{2+}$ influx and efflux through the plasma membrane. The mitochondrial matrix has the ability to sequester $Ca^{2+}$ when free cytosolic $Ca^{2+}$ rises above a set point (Nicholls, 2005). Thus, mitochondria are recognized as one of the sub-cellular organelles participating in regulation of the intracellular $Ca^{2+}$ homeostasis. Mitochondria are dynamic organelles that interact with the plasma membrane and the endoplasmic reticulum (ER) (Boncompagni et al., 2009; Eisner et al., 2013), and contribute to the recycling of $Ca^{2+}$ back to the vicinal ER (Arnaudeau et al., 2001; Frieden et al., 2005). While intracellular $Ca^{2+}$ signaling controls mitochondrial motility, distribution and function (Yi et al., 2004), reciprocally, mitochondria also modulates spatial and temporal intracellular $Ca^{2+}$ levels.

*Corresponding author (email: jzhou@kcumb.edu)

Skeletal muscle contraction needs both $Ca^{2+}$ and ATP. Thus, muscle physiology largely depends on two intracellular organelles: the sarcoplasmic reticulum (SR) for $Ca^{2+}$ storage and release (Franzini-Armstrong and Jorgensen, 1994), and mitochondria for ATP synthesis (Russell et al., 2014). In non-muscle cells, the functional and physical coupling between ER and mitochondria is attributed to the inter-organelle tether proteins called mitofusion at the juxtaposition between the ER and mitochondria (de Brito and Scorrano, 2008). This type of structure was also found in skeletal muscle cells in which a tether like protein connects the SR and mitochondria (Boncompagni et al., 2009; Pietrangelo et al., 2015). These pivotal findings have heightened the role of mitochondria as a key player in the dynamic regulation of physiological $Ca^{2+}$ signaling in skeletal muscle. Although it is believed that there is resemblance of mitochondrial structure and function among all cell types, the way by which mitochondrial $Ca^{2+}$ uptake regulating intracellular $Ca^{2+}$ signaling has specific features in skeletal muscle. Mitochondria in muscle cells face rapid changes of intracellular $Ca^{2+}$ levels during contraction. Whether mitochondria $Ca^{2+}$ uptake modifies $Ca^{2+}$ signaling during excitation-contraction coupling has been a fundamental question in muscle physiology (O'Rourke and Blatter, 2009; Rossi et al., 2009). In order to answer this fundamental question, effort has been made to evaluate mitochondrial $Ca^{2+}$ uptake in skeletal muscle under various physiological conditions. Characterization of mitochondrial $Ca^{2+}$ uptake is a key step to understand the role of mitochondria in muscle physiology and diseases. This review focuses on characterization of mitochondrial $Ca^{2+}$ uptake in skeletal muscle and its significance in skeletal muscle physiology and diseases.

## MITOCHONDRIAL $Ca^{2+}$ UPTAKE REGULATES ENERGY PRODUCTION IN SKELETAL MUSCLE

$Ca^{2+}$ is a critical messenger not only for muscle contraction, but also for promoting mitochondrial ATP production. In mammalian cells, $Ca^{2+}$ is a key regulator of ATP production (Griffiths and Rutter, 2009). Four important mitochondrial dehydrogenase involved in the direct supply of NADH (reduced nicotinamide adenine dinucleotide) and FADH (reduced flavin adenine dinucleotide) for ATP production were found to be regulated by $Ca^{2+}$ inside mitochondria (Denton, 2009). A transient increase of free $Ca^{2+}$ concentration is required to stimulate electron transport chain (ETC) of mitochondria in cardiac cells (Gueguen et al., 2005; Territo et al., 2000). The role of mitochondrial $Ca^{2+}$ uptake in cardiac muscle energy metabolism has been widely studied (Balaban, 2002; Brookes et al., 2004).

In skeletal muscle, ATP demand increases ~100 times during rapid muscle contraction. Such high demand of ATP cannot be fulfilled by the finite amount ATP normally stored inside the skeletal muscle. Muscle contraction requires fast and sustained ATP production, which is fulfilled primarily by mitochondria (Porter and Wall, 2012). As such, skeletal muscle is known to be a tissue of high energy demand with mitochondria occupying 10%–15% of the fiber volume and densely packed within muscle cells (Eisenberg, 1983). In skeletal muscle, mitochondria are located largely within the I-bands, surrounding the SR network (Eisenberg, 1983). Importantly, mitochondria are found to be linked to the SR in skeletal muscle by developmentally regulated tethering structures (Boncompagni et al., 2009; Pietrangelo et al., 2015). This intimate juxtaposition of the SR and mitochondria, together with the ability of mitochondria to take up $Ca^{2+}$ from their surroundings, allows the movement of $Ca^{2+}$ between these organellar systems (Bianchi et al., 2004; Csordas and Hajnoczky, 2009; Rizzuto and Pozzan, 2006; Santo-Domingo and Demaurex, 2010). These movements are believed to help tailor mitochondrial metabolism and ATP synthesis to the demand of muscle contraction. Early studies of intact skeletal muscle observed an increase in NADH/$NAD^+$ during the transition from resting to working status, suggesting that an enhanced intracellular $Ca^{2+}$ level promotes mitochondrial metabolism in skeletal muscle (Duboc et al., 1988; Kunz, 2001; Sahlin, 1985). Later, using isolated mitochondria derived from skeletal muscle, Kavanagh et al. confirmed that an elevation in mitochondrial $Ca^{2+}$ was able to stimulate oxidative phosphorylation (Kavanagh et al., 2000). As discussed in the review article by Rossi et al., mitochondrial $Ca^{2+}$ uptake should assist with stimulation of aerobic ATP production in order to balance increased ATP consumption associated with cross bridge cycling and SERCA-mediated $Ca^{2+}$ sequestration during muscle contraction (Rossi et al., 2009).

## EVALUATION OF MITOCHONDRIAL $Ca^{2+}$ UPTAKE IN SKELETAL MUSCLE

In order to understand the role of mitochondrial $Ca^{2+}$ uptake in skeletal muscle physiology, it is vital to evaluate the amount and the kinetics of mitochondrial $Ca^{2+}$ uptake in skeletal muscle cells under physiological conditions. The early studies on mitochondrial $Ca^{2+}$ uptake were performed on isolated mitochondria (Deluca and Engstrom, 1961; Mraz, 1962). These studies showed that isolated mitochondria from rat kidney were able to take up 60% of $Ca^{2+}$ from the surrounding medium (Deluca and Engstrom, 1961). The kinetics of mitochondrial $Ca^{2+}$ uptake was well documented in the isolated mitochondria from the liver and heart (Carafoli and Crompton, 1978; McMillin-Wood et al., 1980). Sembrowich et al. was the first to explore the $Ca^{2+}$ uptake by mitochondria derived from different types of skeletal muscle both from rats and rabbits (Sembrowich et al., 1985). Using direct patch-clamp recording on the inner mi-tochondrial membrane, Fieni et al. recorded the mitochondrial $Ca^{2+}$ uptake activity in mitoplasts isolated from mitochondria of different types of tissue including skeletal muscle (Fieni et al., 2012). These *in vitro* studies also suggested a potential influence of mitochondrial $Ca^{2+}$ uptake on cytosolic $Ca^{2+}$ signaling during muscle contraction. However, such conclusion needs validation from *in vivo* studies. Specifically, it requires characterization of mitochondrial $Ca^{2+}$ uptake in intact muscle cells under physiological conditions.

There are a few probes available to monitor $Ca^{2+}$ fluxes into and out of mitochondria in live cells. The commercially available fluorescent dyerhod-2 has been widely used in investigating mitochondrial $Ca^{2+}$ handling in cultured cells because the acetoxymethyl (AM) ester of rhod-2 (Rhod-2-AM) preferentially targets mitochondria (see review (Pozzan and Rudolf, 2009)). Rhod-2 has been used to measure mitochondrial $Ca^{2+}$ uptake in cultured skeletal muscle myotubes under electric stimulation (Eisner et al., 2010). The shortcoming is that Rhod-2 is not a ratiometric dye (Fonteriz et al., 2010). The uneven distributions of the dye among individual mitochondria can also cause problems for quantification of mitochondrial $Ca^{2+}$ concentration changes based on fluorescence intensity (Lakin-Thomas and Brand, 1987). Rhod-2 has also been used to monitor mitochondrial $Ca^{2+}$ uptake in intact skeletal muscle fibers following repeated tetanic stimulation (Ainbinder et al., 2015; Bruton et al., 2003). However, the specific targeting of Rhod-2-AM to mitochondria in intact muscle fibers was challenging. To avoid the Rhod-2 signals from outside mitochondria, Shkryl and Shirokova recorded mitochondrial $Ca^{2+}$ uptake during caffeine-induced $Ca^{2+}$ release in permeabilized rat skeletal muscle fibers (Shkryl and Shirokova, 2006). In this case, cell membrane permeabilization of the muscle fibers allowed the non-targeted Rhod-2 dye to leak out of the cytosol. However, since muscle fibers with permeabilized membrane no longer respond to physiological stimulations (i.e. membrane depolarization), the condition employed in such a study is not suitable for quantitative and specific evaluation of mitochondrial $Ca^{2+}$ uptake in intact skeletal muscle cells under physiological conditions.

Due to various limitations, quantitative measurement of mitochondrial $Ca^{2+}$ uptake in skeletal muscle remains to be challenging. GFP and other functionally similar fluorescent proteins have modernized the research in cell biology (Tsien, 1998). Owing to mutations and variations in gene sequences, genetically encoded fluorescent proteins have been developed as $Ca^{2+}$ biosensors with varying properties including differences in fluorescence spectra, $Ca^{2+}$ binding affinities and kinetics as well as those that change spectral properties upon binding to calcium (Palmer et al., 2006). The rapid growth of molecular biology techniques also allows the genetically encoded $Ca^{2+}$ biosensors to target to specific sub-cellular organelles such as mitochondria (Pozzan and Rudolf, 2009). Thus, organelle-targeted ratiometric $Ca^{2+}$ biosensors has become a better choice for characterization of mitochondrial $Ca^{2+}$ uptake in skeletal muscle under physiological conditions. Using a mitochondrial targeted biosensor (2mtYC2), Rudolf et al. demonstrated that a single twitch could cause measurable dynamic changes in mitochondrial $Ca^{2+}$ levels in live skeletal muscle fibers. However, they also noted some limitations of 2mtYC2 for mitochondrial $Ca^{2+}$ measurement in muscle cells, for instance, YC2 had a small dynamic range with an increase of the emission ratio <26% in the cytosol and <14% in mitochondria during muscle contraction (Rudolf et al., 2004). Subsequently, Palmer et al. developed a new version of mitochondrial targeted $Ca^{2+}$ biosensor, 4mtD3cpv, which has a dynamic ratio range of 5.1 (Palmer et al., 2006). Upon testing 4mtD3cpv on live skeletal muscle fibers under voltage-clamp conditions, Zhou et al. found that while 4mtD3cpv showed a significant improvement in monitoring mitochondrial $Ca^{2+}$ levels in live muscle fibers with an increased dynamic ratio range, the kinetics of the detected signal set some limitations for quantitatively calculating the changes of the mitochondrial $Ca^{2+}$ level (Zhou et al., 2008). As an alternative, YC3.6, another $Ca^{2+}$ biosensor constructed by Nagai and colleagues (Nagai et al., 2004), with a dynamic ratio range of 5.6 and apparent $K_d$ of 0.25 µmol L$^{-1}$ was later tested by Yi et al. in live skeletal muscle fibers (Yi et al., 2011). By introducing a mitochondrial targeting sequence (Wang et al., 2008) at the 5′-end of YC3.6 cDNA, they developed a mitochondrial targeting $Ca^{2+}$ biosensor, mt11-YC3.6. The highly specific mitochondrial expression of mt11-YC3.6 and the simple kinetics of the recorded YC3.6 ratio signal allowed quantitative evaluation of the dynamic changes of free $Ca^{2+}$ levels inside mitochondrial matrix in skeletal muscle fibers in response to a $Ca^{2+}$ release transient induced by cell membrane depolarization under whole-cell voltage clamped conditions. This study shows that at the peak of the voltage-induced $Ca^{2+}$ release, the mitochondrial $Ca^{2+}$ uptake contributes to around 10%–18% of the total $Ca^{2+}$ removal, and the average mitochondrial $Ca^{2+}$ influx is around 4.1±1.0 µmol L$^{-1}$ ms$^{-1}$ (Yi et al., 2011). This study represents the first quantitative characterization of mitochondrial $Ca^{2+}$ uptake and its role in shaping the cytosolic $Ca^{2+}$ signaling in skeletal muscle during excitation-contraction coupling.

## IMPAIRED SKELETAL MUSCLE MITOCHONDRIAL $Ca^{2+}$ SIGNALING IN MUSCLE DISEASES

Mitochondrial $Ca^{2+}$ uptake plays vital roles in life and death of the cell. Impaired mitochondrial $Ca^{2+}$ uptake is observed in various skeletal muscle myopathies and neuromuscular diseases. Defective intracellular $Ca^{2+}$ signaling is associated with degeneration of skeletal muscle cells in aging (Delbono, 2002; Weisleder et al., 2006) and muscular dystrophy (mdx) (De Backer et al., 2002; DiFranco et al., 2008; Han et al., 2006; Hopf et al., 1996; Mallouk et al., 2000; Vandebrouck et al., 2002; Wang et al., 2005). Since the defects usually entail increases in the SR $Ca^{2+}$ release activity and elevated myoplasmic $Ca^{2+}$ levels, which likely affect mitochondrial $Ca^{2+}$ uptake. An early study by Robert et al. directly tested this hypothesis by recording mitochondrial $Ca^{2+}$ uptake in myotubes derived from a Duchenne Muscular Dystrophy *mdx* mouse model. Using mitochondria-targeted $Ca^{2+}$-sensitive photoprotein aequorin, they reported that a larger caffeine-induced $Ca^{2+}$ release from the SR led to an augmented mitochondrial $Ca^{2+}$ uptake in the myotubes derived from the *mdx* mice (Robert et al., 2001). A later study by Shkryl et al. confirmed that the excessive myoplasmic $Ca^{2+}$ was taken up by mitochondria in adult skeletal muscle fibers derived from the *mdx* mouse model during osmotically induced $Ca^{2+}$ release (Shkryl et al., 2009). Moreover, genetic mutations that affect mitochondrial function are often associated with skeletal muscle dysfunction. The mitochondrial myopathy mouse model with disruption of the gene for mitochondrial transcriptor factor A (Tfam) shows remarkably altered mitochondrial morphology in skeletal muscle and reduced muscle force (Wredenberg et al., 2002). A later study on skeletal muscle of this mouse model showed that mitochondria accumulated excessive amount of $Ca^{2+}$ following a repetitive contraction (Aydin et al., 2009). Furthermore, mutations in *RyR1* gene encoding the skeletal muscle isoform of the ryanodine receptor (RyR1) cause malignant hyperthermia (MH) and central core disease (CCD). The MH and CCD mutations lead to altered $Ca^{2+}$ release from the SR. By overexpressing the MH and CCD RyR1 mutant proteins in HEK-293 cells, Brini et al. reported a correlation between the level of cytosolic $Ca^{2+}$ transient and the amount of mitochondrial $Ca^{2+}$ uptake, demonstrating that the MH mutation with enhanced cytosolic $Ca^{2+}$ transients simultaneously leads to enhanced mitochondrial $Ca^{2+}$ uptake (Brini et al., 2005). In addition, the knock-in mice harboring the Y522S RyR1 MH mutation showed defective mitochondrial morphology in skeletal muscle (Durham et al., 2008), indicating that uncontrolled $Ca^{2+}$ release due to the mutation in RyR1 leads to mitochondrial damage. Finally, a study on the skeletal muscle fibers derived from aged mice also showed that the increased $Ca^{2+}$ leakage from the SR led to $Ca^{2+}$ accumulation in mitochondria (Andersson et al., 2011). Altogether, the studies listed above support the concept that an enhanced SR $Ca^{2+}$ release or an elevated myoplasmic $Ca^{2+}$ level promotes mitochondrial $Ca^{2+}$ uptake in various muscle diseases. The enhanced mitochondrial $Ca^{2+}$ uptake could lead to $Ca^{2+}$ overload inside mitochondrial matrix and initiate downstream responses leading to muscle cell degeneration, such as excessive mitochondrial ROS production that disrupts the cellular redox state observed in various types of muscle diseases (Durham et al., 2008; Wang et al., 2005; Weisleder et al., 2006).

In skeletal muscle, the intracellular release and uptake of $Ca^{2+}$ are mainly controlled by the SR, which forms a network that is intimately associated with mitochondria. This close spatial proximity between the SR and mitochondria, together with the ability of mitochondria to take up $Ca^{2+}$, suggests that mitochondria could play an important role in shaping intracellular $Ca^{2+}$ signaling in muscle cells. However, whether mitochondrial $Ca^{2+}$ uptake is large and rapid enough to modulate physiological $Ca^{2+}$ transients in skeletal muscle and whether alterations in mitochondrial $Ca^{2+}$-buffering capacity contribute to muscle dysfunction under pathophysiological conditions are fundamental questions for understanding muscle degeneration in various diseases. A direct evidence of mitochondrial regulation on the SR $Ca^{2+}$ release activity in live skeletal muscle cells was obtained from the study on an amyotrophic lateral sclerosis (ALS) mouse model (G93A) with transgenic overexpression of the human ALS-associated SOD1$^{G93A}$ mutant (Zhou et al., 2010). The G93A muscle fibers display localized depolarization of mitochondrial inner membrane potential in the fiber segment near the neuromuscular junction. The depolarized mitochondria lose the driving force for $Ca^{2+}$ uptake, which impairs mitochondrial $Ca^{2+}$ buffering capacity. The fiber segments with depolarized mitochondria shows greater osmotic stress-induced $Ca^{2+}$ release activity, which can include propagating $Ca^{2+}$ waves. Those $Ca^{2+}$ waves are confined to regions of depolarized mitochondria and stop propagating shortly upon entering the regions of normal, polarized mitochondria. Uncoupling of mitochondrial membrane potential with FCCP or inhibition of mitochondrial $Ca^{2+}$ uptake by Ru360 also led to cell-wide propagation of such $Ca^{2+}$ release events. These data reveals that mitochondrial $Ca^{2+}$ uptake is large and rapid enough to shape cytosolic $Ca^{2+}$ signaling in skeletal muscle under physiological conditions.

The ALS muscle fibers provide a unique opportunity to characterize the mitochondrial $Ca^{2+}$ uptake under physiological conditions. The localized mitochondrial defect in the ALS muscle fibers allows for examination of mitochondrial contribution to $Ca^{2+}$ removal during excitation-contraction coupling by comparing $Ca^{2+}$ transients in regions with normal and depolarized mitochondria in the same muscle fiber. Using whole cell voltage-clamp technique, Yi et al. showed that $Ca^{2+}$ transients elicited by membrane depolarization in the fiber segment with depolarized mitochondria displayed increased amplitude of ~10%. Using the mitochondria-targeted $Ca^{2+}$ biosensor (mt11-YC3.6) expressed in ALS muscle fibers, these authors recorded the dynamic change of mitochondrial free $Ca^{2+}$ levels during voltage-induced SR $Ca^{2+}$ release and detected a reduced $Ca^{2+}$ uptake by mitochondria in the fiber segment with depolarized mitochondria, which mirrored the elevated $Ca^{2+}$ transients in the cytosol in the same region (Yi et al., 2011). This study provides a direct demonstration of the importance of mitochondrial $Ca^{2+}$ uptake in shaping cytosolic $Ca^{2+}$ signaling in skeletal muscle during excitation-contraction coupling and suggests that the reduced $Ca^{2+}$ buffering capacity of mitochondria likely contributes to muscle degeneration in ALS.

Although, it was well known that mitochondria from all cell types were able to take up $Ca^{2+}$ and that the channel or transport responsible for mitochondrial $Ca^{2+}$ uptake was defined as mitochondrial $Ca^{2+}$ uniporter (MCU), the molecular identity of the putative MCU had remained mysterious for decades (Carafoli, 2014; Drago et al., 2011; Starkov, 2010). It was not until 2011 when two research groups independently identified the gene that encodes MCU, a transmembrane protein located to the inner mitochondrial membrane (Baughman et al., 2011; De Stefani et al., 2011). This new progress has further advanced the investigation of the role of mitochondrial $Ca^{2+}$ uptake in skeletal muscle health and diseases. Pan et al. generated a global knockout mouse model ($MCU^{-/-}$). The $MCU^{-/-}$ mice survived well with a smaller body size, but showed impaired skeletal muscle performance along with absence of mitochondrial $Ca^{2+}$ uptake in isolated skeletal muscle mitochondria, indicating that mitochondrial $Ca^{2+}$ uptake plays an important role in skeletal muscle development and performance (Pan et al., 2013). Recently, direct evidence of MCU-dependent mitochondrial $Ca^{2+}$ uptake in protecting denervation-induced skeletal muscle atrophy was provided by Mammucari et al. and Chemello et al., in which, the authors have shown that virus-mediated overexpression or silencing of MCU had significant impact on skeletal muscle atrophy through regulation expression of genes involved in hypertrophic pathways in skeletal muscle (Chemello et al., 2015; Mammucari et al., 2015). Although the identified pore-forming molecule of MCU is a highly selective $Ca^{2+}$ channel, other auxiliary subunits participate forming the mitochondrial $Ca^{2+}$ uniportor complex (De Stefani et al., 2016; Jhun et al., 2016; Kamer and Mootha, 2015). The identification of loss-of function mutations in MICU1, a regulator of MCU (Csordas et al., 2013; Perocchi et al., 2010) in patients with proximal muscle myopathy (Logan et al., 2014) indicates the complexity of MCU in skeletal muscle and its role in normal muscle function. However, the precise physiological role and the molecular structure of the mitochondrial $Ca^{2+}$ uniporter complex in skeletal muscle still has more to be determined.

# SUMMARY

Mitochondrial $Ca^{2+}$ uptake is a double-edged sword for muscle function. While the $Ca^{2+}$ influx into mitochondria is required for promoting ATP synthesis, excessive $Ca^{2+}$ accumulation in mitochondria initiates a series of molecular malfunctions leading to mitochondrial damage and cell death. Under diseased conditions, such as muscular dystrophy, gene-mutation related myopathies and aging, enhanced SR $Ca^{2+}$ release activity overloads mitochondria with $Ca^{2+}$, leading to mitochondrial dysfunction and muscle cell degeneration. In those cases, mitochondrial damage seems to be a consequence of extensive elevation of cytosolic $Ca^{2+}$ levels. In ALS G93A skeletal muscle, the mitochondrial membrane potential is depolarized, which leads to a reduced $Ca^{2+}$ buffering capacity of mitochondria. This reduced mitochondrial $Ca^{2+}$ uptake further overloads those polarized mitochondria with $Ca^{2+}$ and causes further mitochondrial damage in the same cell. In this case, the compromised mitochondrial $Ca^{2+}$ uptake is a leading cause of the disrupted intracellular $Ca^{2+}$ signaling that initiates muscle cell degeneration. In summary, any dysregulation in the amount and kinetics of mitochondrial $Ca^{2+}$ uptake will cause mitochondrial dysfunction and abnormal intracellular $Ca^{2+}$ signaling that leads to muscle cell degeneration. It is predicted that identification of molecular basis associated with mitochondrial $Ca^{2+}$ uptake will further advance the understanding of the role of mitochondrial $Ca^{2+}$ uptake in skeletal muscle health and diseases.

**Compliance and ethics** *The author(s) declare that they have no conflict of interest.*

**Acknowledgements** *This work was supported by National Institute of Arthritis and Musculoskeletal and Skin Diseases (NIAMS)/National Institutes of Health (NIH) Grant (R01 AR057404) to Jingsong Zhou. Funder plays no role for this study, in design, data collecting, data analysis and interpretation, and manuscript writing.*